# When is using AI the rational choice?
# The importance of counterfactuals in AI deployment decisions[1]

Paul Lehner and Elinor Yeo
The MITRE Corporation

April 2025

## Abstract

Decisions to deploy AI capabilities are often driven by counterfactuals - a comparison of decisions made using AI to decisions that *would have been* made if the AI were not used. *Counterfactual misses*, which are poor decisions that are attributable to using AI, may have disproportionate disutility to AI deployment decision makers. *Counterfactual hits*, which are good decisions attributable to AI usage, may provide little benefit beyond the benefit of better decisions. This paper explores how to include counterfactual outcomes into usage decision expected utility assessments. Several properties emerge when counterfactuals are explicitly included. First, there are many contexts where the expected utility of AI usage is positive for intended beneficiaries and strongly negative for stakeholders and deployment decision makers. Second, high levels of complementarity, where differing AI and user assessments are merged beneficially, often leads to substantial disutility for stakeholders. Third, apparently small changes in how users interact with an AI capability can substantially impact stakeholder utility. Fourth, cognitive biases such as expert overconfidence and hindsight bias exacerbate the perceived frequency of costly counterfactual misses. The expected utility assessment approach presented here is intended to help AI developers and deployment decision makers to navigate the subtle but substantial impact of counterfactuals so as to better ensure that beneficial AI capabilities are used.

---

[1] This material is based upon work supported by the AI Research Institutes Program funded by the National Science Foundation under AI Institute for Societal Decision Making (AI-SDM), Award No. 2229881.

# When is using AI the rational choice?
# The importance of counterfactuals in usage decisions

**Introduction**

Consider the following.

> In September 2023 a report was released comparing the safety record of GM Cruise self-driving taxis in San Francisco to human-driven ride-hail taxis that provided a similar service (Zhang, 2023). The study showed that the self-driving taxis had substantially fewer accidents and injuries, with no deaths. Particularly striking was the 94% reduction in accidents where the taxi was faulted for causing the accident.
>
> About a week after this report was released there was an accident that led to headlines such as "*Freak accident in San Francisco traps pedestrian under robotaxi*" (Washington Post, 10/3/2023). A human-driven vehicle hit a pedestrian. The pedestrian was pushed into a Cruise self-driving taxi which initially stopped but then proceeded to pull to the side of the road dragging the pedestrian underneath for about 20 feet. The pedestrian sustained critical injuries from this sequence of events.
>
> Importantly, nothing about this occurrence belied the statistics on the relative safety of robotaxis. Nevertheless this one event led California to pull the license for Cruise taxis and GM to pull the taxis off the road everywhere (Howland and Chrisner, 2023). About a year after this event, GM decided to completely discontinue its investment in self-driving taxis.

This was a "freak" accident because, counterfactually, a human driver would not have made the mistake of dragging the pedestrian.

This story illustrates a case where the considerable net benefit of using AI was overridden by an episodic *counterfactual miss* – a negative event that would not have occurred if a human had made the decision. While some may debate the validity of the statistical safety analysis, we note that overall safety rates did not appear to be a major consideration in the usage decision to pull all Cruise taxis off the street.

This paper presents an approach to expected utility modeling of AI deployment decisions that include exposure to episodic counterfactual outcomes. Using this rational choice approach we show that from the perspective of a deployment/usage decision maker,
1. Using an AI capability may have negative expected utility for a usage decision maker even when there is clearly positive utility for intended beneficiaries
2. The conditions under which an AI capability provides the greatest net benefit are sometimes the same conditions where disutility to usage decision makers is at its worst.

3. Small changes in how a user interacts with an AI may substantially impact stakeholder and deployment decision maker expected utility. Adoption and enduring usage is very dependent on the structure of user/AI interactions.

**Approach**

Begin with the conceptual framework depicted in Figure 1. Usage/deployment decisions are driven by both beneficiary and stakeholder utilities, where the latter is often driven by societal factors. In the Cruise taxi example, the public was the beneficiary, while GM and the California DMV were both stakeholders and deployment decision makers. Both CA and GM were substantially influenced by societal reactions to the counterfactual miss – the "freak" accident.

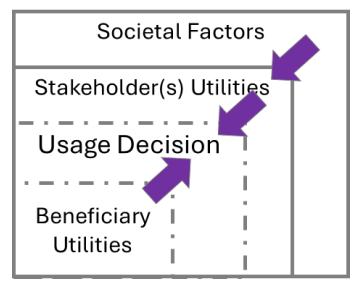

Figure 1: Framework

A second example, which is used herein as a running example, is use of AI to support medical decision making. In particular, clinicians (usually radiologists) using AI image reading to help detect cancer. The beneficiaries are patients. The primary stakeholders are clinicians and medical facilities which house the radiology service (medical practice, hospital, medical lab, etc.). Societal factors include such things as press coverage, medical ethics, malpractice lawsuits, etc. Clinicians are the users. Both clinicians and medical facilities are the usage decision makers who determine if AI image reading should, or should continue to be, deployed.

Numerous other examples will be discussed in the Discussion section.

*Counterfactual outcomes matrix*

Key to an expected utility analysis of AI usage is the *counterfactual outcomes matrix*. Table 1 depicts the elements of such a matrix for binary judgments. The cells in a traditional 2x2 outcome matrix are partitioned into non-counterfactual and counterfactual outcomes.

| GT \ AI+DM,DM-U | AI+DM="T" DM-U="T" | AI+DM="T" DM-U="F" | AI+DM="F" DM-U="T" | AI+DM="F" DM-U="F" |
|---|---|---|---|---|
| T | Non-Counterfactual True Positive (NTP) | Counterfactual True Positive (CTP) | Counterfactual False Negative (CFN) | Non-Counterfactual False Negative (NFN) |
| F | Non-Counterfactual False Positive (NFP) | Counterfactual False Positive (CFP) | Counterfactual True Negative (CTN) | Non-Counterfactual True Negative (NTN) |

Table 1: Counterfactual outcomes matrix for binary judgments

Here
- GT is the ground truth, which is either T or F,
- AI+DM is the combination of the AI and decision maker (the user). AI+DM = "T" means that using the AI the DM concluded "T".
- DM-U is the unaided decision maker. Here DM-U = "T" means that the DM *would have* decided "T" if she had not used the AI.

Consider as an example a counterfactual false negative (CFN). In our radiology example this reads as
1. The patient has cancer (GT=T).
2. Using AI image reading the clinician falsely diagnosed that the patient did not have cancer (AI+DM = "F").
3. If the clinician had not used the AI, she would have correctly diagnosed that the patient had cancer (DM-U = "T").

The cancer diagnosis was missed *because* the AI was used.[2]

By contrast, a Counterfactual True Positive (CTP) is a case where cancer was discovered because AI image reading was used; unaided the clinician would have missed it.

A *counterfactual miss* is either a CFN or CFP, while a *counterfactual hit* is either a CTP or CTN.

An expected utility calculation needs assignment of probabilities and utilities to the cells in this matrix, but there are important and impactful nuances that must be considered. These are described below.

---

[2] The definitional relationship between causality and counterfactuals is nuanced (e.g., Menzies and Beebee, 2024). Sometimes notions of causality are used to define counterfactuals and sometimes counterfactuals are used to define the concept of causality. Within the scope of this paper, a statement that the AI caused a decision is defined by an episodic counterfactual claim that if the AI were not used in a particular instance, then the decision would have been different. A deeper understanding of the relationship between causality and counterfactuals is not needed here.

*Ingredients of EU analyses of usage decisions that include counterfactual outcomes*

This subsection describes the parameters needed for an expected utility analysis of AI usage decisions that includes exposure to episodic counterfactuals. The focus in this section is on the structure of an EU model.

Counterfactual outcome matrix probabilities[3]

The first ingredient is counterfactual outcome probabilities as illustrated in Table 2 below.

|  | AI+DM = "T" DM-U = "T" | AI+DM = "T" DM-U = "F" | AI+DM = "F" DM-U = "T" | AI+DM = "F" DM-U = "F" |
|---|---|---|---|---|
| GT = T | 0.135 | 0.025 | 0.01 | 0.03 |
| GT = F | 0.09 | 0.07 | 0.09 | 0.55 |

Table 2. Illustrative counterfactual outcome probabilities

"Partitioning" this matrix into probabilities for AI-aided decision making (AI+DM) and unaided decision-making (DM-U) yields Table 3.

| 3a | AI+DM="T" | AI+DM="F" |
|---|---|---|
| GT=T | 0.16 | 0.04 |
| GT=F | 0.16 | 0.64 |

| 3b | DM-U="T" | DM-U="F" |
|---|---|---|
| GT=T | 0.145 | 0.055 |
| GT=F | 0.18 | 0.62 |

Table 3. Aided and unaided outcome probabilities
derived from counterfactual outcome probabilities

Sensitivity and specificity are better for AI+DM (both at 0.8) than for DM-U (0.725 and 0.775). The improvements in sensitivity and specificity are entirely due to the frequency difference between CTPs and CFNs, and between CTNs and CFPs.

Outcome utilities

The second ingredient is outcome utilities. *Outcome utilities are utilities attached to outcomes that are independent of the decision process.* For our running example we will use the outcome utilities shown in Table 4.

---

[3] There are long standing and vigorous debates about the meaning of the word "probability", where the differing definitions have strong practical implications (see Lehner and Lasky, 1996 for quick overview). In this paper, it will suffice to interpret "probability" as meaning the relative frequency of future cases satisfying certain conditions, and probability values as estimates of those relative frequencies. A more nuanced understanding of the meaning of the word "probability" is not needed here.

|  | AI+DM = "T" DM-U = "T" | AI+DM = "T" DM-U = "F" | AI+DM = "F" DM-U = "T" | AI+DM = "F" DM-U = "F" |
|---|---|---|---|---|
| GT = T | 2 | 2 | -10 | -10 |
| GT = F | -1 | -1 | 1 | 1 |

Table 4. Illustrative outcome utilities

Here a correct negative diagnosis is treated as a reference point. The patient enters the facility, is correctly diagnosed, the fee is paid, and no follow-up is needed. This receives a baseline utility of 1 for the medical facility. A false positive has a small negative value. On rare occasions false positives lead to negative consequences (e.g., malpractice suit due to unnecessary exploratory surgery) but for our illustration we assert that on average false positives are not a substantial cost to the facility or clinician. By contrast false negatives can be quite costly. Missed or delayed diagnoses are among the primary reasons for malpractice lawsuits (Shaffer, et. al., 2017). While most suits are dismissed the possibility of such suits is reflected in malpractice insurance costs for both clinicians and medical facilities.[4]

Importantly, counterfactual and non-counterfactual outcomes always have the same utilities. This is because these utilities are associated with the combination of decision and outcome, and not how the decision was obtained. Outcome utilities are the same whether or not AI is used.

Counterfactual outcome utilities

Counterfactual outcome utilities are utilities that are tied to the combination of the decision, outcome *and* differences in *how* that decision was made. Table 5 shows illustrative counterfactual utilities for our running example.

|  | AI+DM = "T" DM-U = "T" | AI+DM = "T" DM-U = "F" | AI+DM = "F" DM-U = "T" | AI+DM = "F" DM-U = "F" |
|---|---|---|---|---|
| GT = T | 0 | 5 | -30 | 0 |
| GT = F | 0 | -2 | 5 | 0 |

Table 5. Illustrative counterfactual outcome utilities

For this illustration non-counterfactual outcomes have neither positive nor negative utility. Whether or not the AI is used would have led to exactly the same decision. But counterfactual outcomes can have an outsized effect. Consider in particular the -30 in the CFN cell. This reflects concerns about the following easily imagined sequence of events.
1. There is a missed cancer diagnosis that leads to a negative health outcome.
2. A clinical care review is initiated to try to understand how the diagnosis was missed.
3. The review retrospectively concludes that if the clinician had not used the AI, she would not have made this error ("How could you have missed that?").

---

[4] We appreciate that this oversimplifies utilities and costs in cancer screening, but for purposes of this paper this simple characterization will suffice.

4. Because clinicians have an ethical obligation to make final judgments themselves (Kempt, H., 2021), the clinician may be formally chastised for her lack of medical ethics. She appears to have deferred to the AI.
5. A malpractice suit could soon follow where, given the documented unethical practice, the suit is quickly settled in favor of the plaintiff.

On the plus side are CTPs. In our running example these are cases where the AI led to finding cancer that the clinician would otherwise have missed. Here we give this circumstance an illustrative (and generous) utility of 5 under the belief that positive press and benefits would obtain from documented cases where use of an AI capability led to a possibly lifesaving diagnosis.

Counterfactual discovery probabilities.

For a counterfactual outcome utility to obtain, the counterfactual outcome must be noticed. *There must be a retrospective review that determines how the judgment was made*. In our running example, CFNs are noticed because missed diagnoses routinely spark clinical care reviews. CTPs are rarely noticed because there is little reason to initiate a clinical care review of a correct diagnosis.

Table 6 is a counterfactual discovery probability matrix that estimates the frequency that different counterfactuals will be noticed.

|  | AI+DM = "T" DM-U = "T" | AI+DM = "T" DM-U = "F" | AI+DM = "F" DM-U = "T" | AI+DM = "F" DM-U = "F" |
|---|---|---|---|---|
| GT = T | n/a | .01 | .8 | n/a |
| GT = F | n/a | .1 | .01 | n/a |

Table 6. Illustrative counterfactual discovery probabilities

Here is posited a 80% chance that an AI-attributable missed cancer diagnosis will be documented and only a 1% chance that an AI-attributable correct cancer diagnosis will be documented.

*Expected utility calculation*

Given the above ingredients, the AI usage-EU calculation is

| Usage-EU | = | Outcome-EU | + | Counter-EU |
|---|---|---|---|---|
| | = | $\sum_{ij} (p_{ij} \cdot U_{ij})$ | + | $\sum_{ij} (p_{ij} \cdot d_{ij} \cdot CU_{ij})$ |
| | = | 0.400 | + | -0.24825 |
| | = | | 0.15175 | |

Where $U_{ij}$ is an outcome utility, $CU_{ij}$ is a counterfactual outcome utility, $p_{ij}$ is a counterfactual outcome matrix probability and $d_{ij}$ is a discovery probability. Notice that the contribution of the counterfactual outcomes is negative.

Using the outcome probabilities shown in Table 3b and outcome utilities in Table 4

$$\text{Unaided-EU} = 0.18$$

This gives us

$$\text{Outcome-EU} > \text{Unaided-EU} > \text{Usage-EU}$$
$$0.400 > 0.18 > 0.15175$$

Using the AI more than doubles benefit (outcome utility). But the usage decision maker would see a nearly 16% drop in their EU. Despite clearly better health outcomes, the usage decision maker would be wise to avoid using the AI.

Of course the above analysis is entirely dependent on the illustrative numbers. However, as will be shown below, in contexts where there are discoverable costly counterfactual misses, it can be difficult to obtain usage-EU > unaided-EU.

**Application strategies**

The usage-EU equations described above can be applied in various ways. Four qualitatively different approaches are described below. But first, two items of additional context are needed.

In this paper the phrase "rational choice" reflects the view that expected utility thinking is the axiomatically-derived normative ideal for decision making (Von Neumann and Morgenstern, 1953; Savage, 1954). It is however an unrealistic ideal (Laskey and Lehner, 1994). The best that can be achieved in practice is to approximate this ideal. Importantly, nothing in the axiomatic basis of the ideal entails that the best practical approximation is to build an expected utility model with explicitly declared probabilities and utilities. So, while the application strategies described below are guided by the usage-EU equation, not all of them involve explicit utility and probability elicitation. Indeed, an application strategy that involves subjective probability elicitation is discouraged.

Second, there already exist a variety of frameworks that articulate factors that contribute to AI adoption (e.g., Hameed, et.al., 2023; Kurup and Gupta, 2022; Booyse and Schepers, 2023; Rawashdeh, et.al., 2023). Roughly described these frameworks endeavor to exhaustively enumerate the collection of factors that influence the utility of using an AI capability, which includes variables such as user acceptance, match to organizational processes, maintenance costs, reliability, etc. Nothing in this paper contradicts these frameworks. To our knowledge however none of them explicitly reference counterfactual outcomes. The application strategies presented below focus on counterfactuals. But be aware that counterfactual considerations would only a part of a full scope analysis of AI adoption. In future work we intend to contribute to a comprehensive approach to evaluating the value of potential AI applications that combines traditional and counterfactual considerations.

Four different strategies for applying the AI usage-EU equation are described below: judgment-based estimation, data-driven estimation, structural analysis and simulation-based design. The judgment and data-driven approaches presuppose a specific plan for how DMs will interact with the AI, whereas the structural and simulation-based analyses are intended to help developers to select an effective strategy for user interactions with an AI. The simulation results shown below will illustrate how seemingly small changes in how a user interacts with the AI can lead to dramatic swings in usage-EU for deployment decision makers.

*Judgment-based estimation*

Here judgment-based estimation refers to asking experts to directly estimate probabilities and utilities. In radiology, for example, a medical practice might be interested in using AI image reading and could ask clinicians to provide estimates of both probabilities and utilities.

There is a long and deep literature on eliciting utilities from experts (e.g., Huber, 1974; Morgan, 2014). Often utilities are derived (or at least tested) against forced choice judgments. ("Would you tolerate one CFN to attain three CTPs? If not, how about four CTPs?") These methods begin with clearly stated outcomes and could be applied directly to counterfactual outcomes. There is nothing uniquely challenging about subjective assignment of utilities to counterfactual outcomes.

Uniquely challenging are counterfactual probabilities. Experts would be asked how often they or other expert judgments will be wrong. And when they are wrong, what are the chances that the AI would be right? Given the extensive research literature documenting expert overconfidence in probability judgments and self-assessments (e.g., Tetlock, 2005; Sanchez and Dunning, 2023) it is reasonable to expect that expert estimates will overestimate DM-U accuracy, and they are not likely to see many circumstances where the AI will be correct when they are wrong. Furthermore, other cognitive biases, such as hindsight bias (e.g., Arkes, 2013), would likely cloud expert retrospective self-assessments on individual cases. Where there was an AI error, experts might confidently yet incorrectly believe "they would not have made that mistake".[5]

The directional impact of such cognitive biases in counterfactual probability estimation is depicted in Table 7.

---

[5] Even without AI, hindsight bias is already a known problem in malpractice suits. For example, Berlin (2000) describes a malpractice suit where a radiologist missed a cancer diagnosis. The defense, and his expert radiologist witnesses, argued that the correct reading could only have been made after the location of the cancer was known, and that radiologists who claim they *would have* seen the cancer in the original image were exhibiting hindsight bias. The plaintiff was awarded over $800,000.

| GT \ AI+DM,DM-U | AI+DM="T" DM-U="T" | AI+DM="T" DM-U="F" | AI+DM="F" DM-U="T" | AI+DM="F" DM-U="F" |
|---|---|---|---|---|
| T | Non-Counterfactual True Positive (NTP) ← | Counterfactual True Positive (CTP) | Counterfactual False Negative (CFN) ← | Non-Counterfactual False Negative (NFN) |
| F | Non-Counterfactual False Positive (NFP) → | Counterfactual False Positive (CFP) | Counterfactual True Negative (CTN) → | Non-Counterfactual True Negative (NTN) |

Table 7. Impact of expert overconfidence and hindsight biases on estimates of counterfactual probabilities.

Overall judgment-based estimates are expected to overestimate the frequency of counterfactual misses and to underestimate the frequency of counterfactual hits. In domains where there are substantial costs for counterfactual misses, it is difficult to imagine a decision support scenario where a deployment decision that is based entirely on expert judgment would yield the conclusion usage-EU > unaided-EU.

*Data-driven estimation.*

Here data-driven means empirical data collection where the results of the collection directly extrapolate to assigning parameter values. In radiology this might proceed as follows.
1. Counterfactual utilities: Monetize outcomes and counterfactual outcomes. For medical facilities and insurance companies this is already standard practice. Malpractice insurance costs are based on such data-driven analyses.
2. Counterfactual probabilities:
    a. Collect data on AI and DM-U image reading on common cases. Some data sets already exists (e.g., Leibig, et. al., 2022).
    b. If needed, separately collect data on AI and AI+DM image reading on common cases. AI+DM readings must be with clinicians other than those in 2a.
3. Counterfactual discovery probabilities: Collect historical data on conditions under which a clinical care review is undertaken and documented.

The unusual element in the above data collection is 2b. If AI+DM decisions are automated, such as when AI and DM-U judgments are combined algorithmically, then directly collecting data on AI+DM decisions is unnecessary. Expected AI+DM judgments can be determined without involving human trials. This approach was used by Leibig (2022), Schaffter (2020) and Steyvers (2022); where each study showed how combining AI and DM-U algorithmically would yield better accuracy than either AI or DM-U alone. Also, if the proposed usage has DMs making an initial judgement before invoking the AI, then 2a and 2b can be merged into a single data collection with the same clinicians.

Frequently, however proposed usage is one where the AI generates an initial recommendation and explanation, the DM reviews the AI output, and then the DM makes a final determination. The recent surge of interest in explainable-AI (Dwivedi, 2023), especially as applied to decision support (Yuhan, 2023), is largely founded on this assumed approach to using an AI. With this type of usage it is impractical to collect both DM-U and AI+DM judgements from the same individuals. Imagine a clinician finding evidence of cancer under the AI+DM condition and then being asked to forget everything she saw to make an independent DM-U judgment. Clearly impractical.

Nevertheless, despite some methodological challenges, a data-driven usage-EU analysis can be developed.

*Structural analysis*

The core of an EU analysis that includes counterfactuals is the 2x2 submatrix of counterfactual outcomes. Any usage-EU analysis should begin with an articulation of the possible counterfactuals and whether they are discoverable. Sometimes just that articulation is sufficient to draw strong conclusions about usage-EU. To illustrate this, consider the COMPAS (Correctional Offender Management Profiling for Alternative Sanctions) algorithm that that generates recidivism forecasts (McCay, 2019). Though it is widely used, there are concerns about systematic prediction errors (Chaio, 2019; Rudin et.al., 2020; Engle, 2024) and there are efforts to develop algorithms that are more accurate. Assume an algorithm, call it CP+, replaces COMPAS. The set of counterfactual outcomes comparing the new and old algorithm as applied to parole decisions are shown in Table 8.

|   | CP+ = "P"<br>CP = "P" | CP+ = "P"<br>CP = "NP" | CP+ = "NP"<br>CP = "P" | CP+ = "NP"<br>CP = "NP" |
|---|---|---|---|---|
| SC = T |   | CP+ → parole,<br>CP → no parole<br>Serious crime (would be) committed | CP+ → no parole,<br>CP → parole<br>Serious crime (would be) committed |   |
| SC = F |   | CP+ → parole,<br>CP → no parole<br>No serious crime (would be) committed | CP+ → no parole,<br>CP → parole<br>No serious crime (would be) committed |   |

Table 8. Possible counterfactual outcomes if COMPAS is replaced with a new algorithm.

Here SC=T means an individual will commit a serious crime if paroled, and SC=F that they will not. CP+="P" means that the new algorithm generated a recidivism risk prediction that suggested that parole was the correct decision and the person was paroled. CP="NP" means that the COMPAS recidivism prediction would suggest denying parole. Table 9 shows the discovery profile if CP+ fully replaces COMPAS.

|   | CP+ = "P"<br>CP = "NP" | CP+ = "NP"<br>CP = "P" |
|---|---|---|
| SC = T | Serious crime leads to external request for review where CP+ is (correctly) blamed for choice | No parole, so no discovery |
| SC = F | No serious crime so no discovery | No parole, so no discovery |

Table 9. Discoverability of counterfactual outcomes if COMPAS is replaced with a new algorithm.

As can be seen, the only discoverable counterfactuals are episodes where the new algorithm led to parole for someone who commits a serious crime when the original algorithm would have recommended denying parole. It seems likely that a news service that covers serious crimes, or parties with allegiance to the original algorithm, will discover one such instance. And once the first one is discovered, and reported in the media, others will be found. As a consequence, one should expect that every news story about the new algorithm will be substantially negative, possibly leading to rejection of the new algorithm. And that this result will obtain even if the new algorithm is considerably measurably more accurate than the original.

The problem here is not the new algorithm, but in how it is deployed. Specifically, a *use pattern* (how the AI is used) where the new algorithm entirely replaces the old could sow the seeds of rejection of the new algorithm. Imagine instead a very different use pattern – where the new algorithm is fielded alongside the existing algorithm. Here judges and parole boards receive recommendations from both algorithms, which should result in the discovery pattern shown in Table 10.

|  | CP+ = "P" <br> CP = "NP" | CP+ = "NP" <br> CP = "P" |
|---|---|---|
| SC = T | Serious crime leads to external request for review where CP+ is (correctly) blamed for choice | Serious crime leads to external request for review where CP is (correctly) blamed for choice |
| SC = F | No serious crime so no discovery | No serious crime so no discovery |

Table 10. Discoverability of counterfactual outcomes if COMPAS and new algorithm are fielded together.

Cases where the new algorithm would have prevented parole for a future offender are now also discoverable, undercutting the one-sided sequence of bad news stories for the new algorithm.

This example demonstrates how a change in the use pattern can substantially impact the prognosis for adoption. The section below will show that this is common.

*Simulation based design*

Described below is a simulation model for binary decisions that examine usage-EU under diverse conditions. This simulation model was developed with several objectives in mind:
1. Provide design guidance to AI tool developers to increase the chances of actual and enduring usage of any AI capability they are developing.
2. Generate forecasts on the evolution of usage of an AI tool that is deployed, including predictions as to whether usage will discontinue.
3. Support some policy tradeoff analyses; especially in cases where current policies inhibit usage of measurably beneficial AI tools.

We note that the primary interest here is modeling usage of AI tools that *should* be used; AI tools where it is already well established that outcome-EU > unaided-EU.

Within the simulation a "scenario" requires setting 42 different parameters and each simulation run allows users to compare five different scenarios.  This allows simulation users to examine how parameter changes impact expected utilities.

Rather than tediously describe each of the 42 parameters (see the appendix for that), presented below are some illustrative yet informative simulation results.  For each result the key parameter settings are described along with an explanation of key results that arguably generalize.   An addendum to this paper will be made available that allows readers to independently replicate these findings and to try different parameter settings.

Sim1: Using AI to automate decision making

Figure 2 shows results for a hypothetical AI capability where the following conditions apply.  Parameters are highlighted.
- **AI accuracy** is set to be a little better than **DM-U accuracy**.
- **Use pattern** is set to having the DM automatically accept every AI recommendation.  This is use-pattern 1 or UP1.
- **Utilities and discovery probabilities** are set to the values shown in Tables 4-6 above,
- **Algorithm complementarity** between the AI and DM-U is varied across the five scenarios.  Algorithm complementarity is here defined as the extent to which AI and DM-U make different judgments on cases, while keeping AI and DM-U accuracy constant.[6]
- The many other parameters are all set at moderate levels.

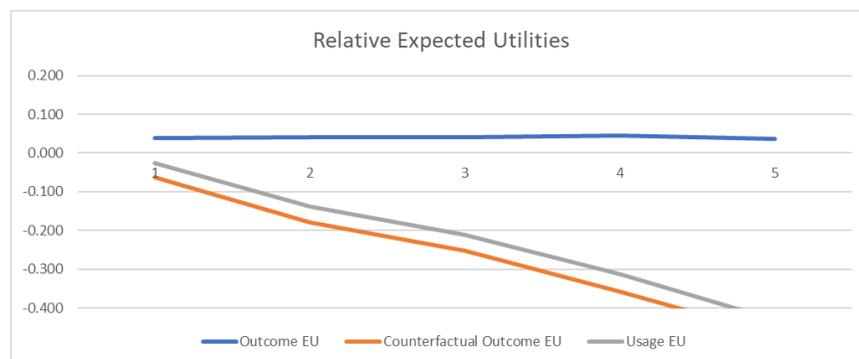

Figure 2.  Expected utilities for increasing levels of complementarity
when DM routinely accepts AI recommendations without review.

Here, the X-axis reflects increasing levels of complementarity.  The Y-axis are relative expected utilities, where 0.0 is the reference Unaided-EU.

---

[6] There is a rich and growing research literature on human-AI complementarity (e.g., Hemmer, 2024). Algorithm complementarity is just one form of human-AI complementarity.

Notice first that outcome-EU is positive and flat across the different complementarity levels. This occurs because the AI is more accurate than DM-U, and the AI and DM do not interact to generate a final decision. In the context of our running example this would mean that patients are generally better off if clinicians and medical practices use AI to automate cancer screening, even though the AI may make some judgments that are different than clinician judgments.

Counter-EU is negative and degrades quickly with increasing complementarity. This result is driven almost entirely by the discoverable CFNs. Other counterfactual outcomes have little impact because they either have small utilities (positive or negative), are unlikely to be discovered, or both. It only takes a few CFNs to push counter-EU very negative, and increasing complementarity increases the *absolute* frequency of CFNs.

At the lowest level of algorithm complementarity, when AI and DM-U make largely redundant judgments, usage-EU is slightly negative.[7] As complementarity increases, usage-EU becomes increasingly negative.

Finally, we note that usage-EU as described above is an average *per-case* calculation. Productivity gains that may be attributable to AI usage are not directly considered. However, if usage-EU is positive, then an ability to process more cases should result in greater gains and if usage-EU is negative, then increasing productivity should magnify the negative result.

From the perspective of our running example, given these parameter settings, Sim1 suggests that automated AI image reading should not be deployed despite clear benefit to patients.

Sim2: Symmetric triage.

A use pattern is a standard sequence of interactions with the AI. UP1 above was to simply accept the AI recommendation. Introduced here is UP2 where AI is used for triage. Specifically,
- If the AI confidently concludes "T" then the DM accepts that conclusion.
- If the AI confidently concludes "F" then the DM accepts that conclusion.
- If the AI presents a conclusion without confidence, the DM ignores the AI entirely and proceeds to address the case herself.

Figure 3 shows simulation results where all settings are exactly the same as in Sim1, except that the AI is used for triage. The confidence thresholds for automatic acceptance are symmetric for positive and negative readings and approximately 65% of the cases exceed the confidence thresholds. Consequently, the AI reading was automatically accepted on around 65% of the cases and DM-U worked about 35%.

---

[7] In the simulation, algorithm complementarity only applies to the portion of evidence available to both the AI and DM. The AI or DM may see additional evidence. So even at the most redundant settings, there is still some difference between AI and DM-U judgments.

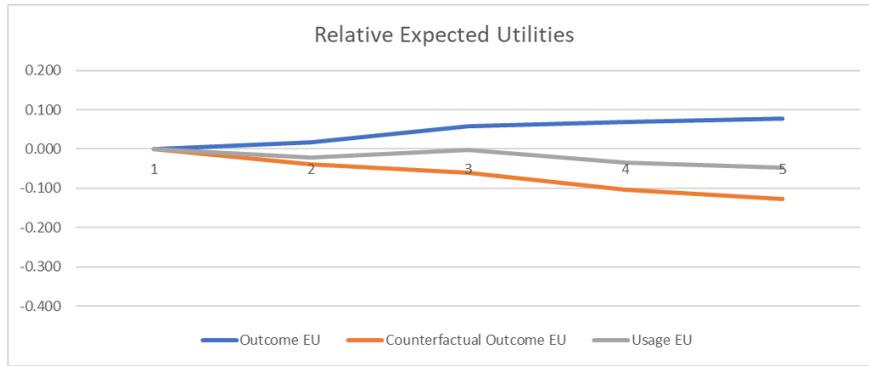

Figure 3. Sample run with UP2: Accept AI if AI confident else solve unaided

There are several things to take note of here.

First, the pattern of expected utility is vastly different than in the previous simulation run. The only difference between Sim1 and Sim2 is the use pattern. This is a common result. Changes in use pattern will, by itself, substantially impact outcome-EU and usage-EU.

Second, outcome-EU improves with increasing complementarity. At the highest level of complementarity the combination of AI and DM yields outcome-EU that is better than either alone.[8] This is not an anomaly. At the lowest complementarity level AI and DM-U decisions are largely the same. There is little room for improvement. At high levels, AI and DM-U often disagree. Either the AI or the DM is correct, and the other is wrong. A good use pattern gravitates the combined AI+DM decision toward whichever is correct. Leibig et. al. (2022) in fact empirically demonstrated exactly this result, with a similar use pattern, in radiology.

Third, counter-EU and usage-EU become increasingly negative as complementarity increases. Furthermore, usage-EU is definitively negative at high complementarity levels. Again this is not an anomaly. Increasing complementarity will increase the frequency of both counterfactual hits and counterfactual misses. But the counterfactual misses are very costly, so the increase in the frequency of counterfactual misses easily outweighs the benefits of better decisions.

Finally, combining the above observations, in Sim2 we find that *the condition where outcome-EU is maximally positive is precisely the condition where usage-EU is maximally negative*. This is not an anomaly. It is a characteristic of many use patterns.

---

[8] This is calculated as follows. Unaided-EU is 0 because that is how 0 is defined. Per the results in Sim1 outcome-EU for the AI varies between 0.037 and 0.045, with an average value of 0.041. In Sim2, Outcome-EU at the highest complementarity level is 0.078. Comparing 0.078 to 0.041 suggests that, at the highest complementarity level, AI+DM yields approximately 90% improvement on outcome-EU compared to the AI alone.

Sim3: Asymmetric triage.

If the consequences of a few CFNs are severe then it makes sense to try to avoid them entirely. One approach to doing this is to modify the UP2 confidence threshold for negative readings to ensure that the AI never generates a confident negative reading.

Figure 4 shows results when a confidence threshold is reset to ensure that there are no confident negative readings. All other parameters are exactly the same as they were in Sim2.

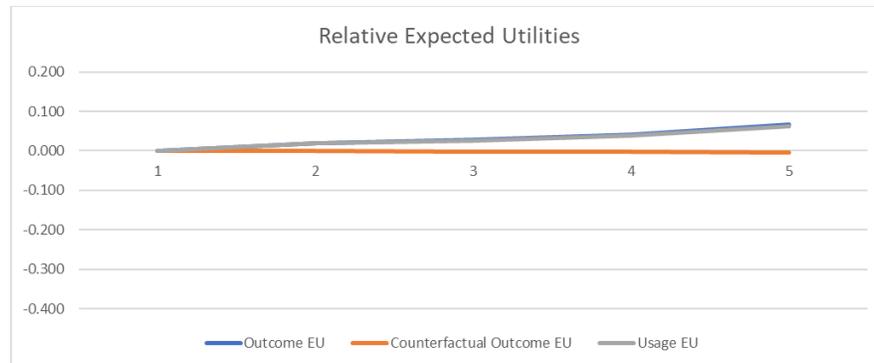

Figure 4: Sample run with UP2, where all negative AI readings are ignored.

Success! Counter-EU is roughly 0 at all complementarity levels, outcome-EU increases with complementarity as does usage-EU. Furthermore at the highest complementarity level outcome-EU = 0.066 which is still 61% higher than the AI alone outcome-EU of 0.041.

In the context of our running example this implies that patients, clinicians and medical facilities all will benefit from using AI image reading in this way. The rational choice is to use it. And if it is used, usage will continue. We refer to this as a ***robust use pattern*** because all stakeholders benefit.

Sim4: Asymmetric triage with Explainable-AI

The simulations described above make no use of explainable-AI. Either the AI or DM-U makes the judgment, and triage criteria are used to decide which. This contrasts with much of the AI literature where it is often assumed that users will consider AI recommendations and explanations before making their final judgments.

UP3 adds explainable-AI to UP2. Specifically,
- If the AI confidently concludes "T" (cancer) then the user accepts that conclusion.
- If the AI confidently concludes "F" (no cancer) then the user accepts that conclusion.
- If the AI presents a conclusion without confidence, the user considers the AI recommendation and explanation before making a final judgment.

In the simulation the impact explainable-AI is modelled with two sets of parameters.

- The first set characterizes the extent to which AI outputs move DM judgments in the direction of the AI recommendations. If explainable-AI is intended to justify an AI recommendation, then presumably it will influence users to accept its recommendation, even when the AI recommendation is wrong.
- The second set characterizes the extent to which explainable-AI helps users move their judgment in the correct direction whether or not the AI is correct. On occasion the AI may be wrong, explainable-AI justifies the wrong answer, but the DM looks at the explanation content and sees something that leads her toward the correct conclusion.

Figure 5 shows results where all parameters are exactly the same as Sim3, except that the user considers the AI recommendations and explanations before making a final judgment.

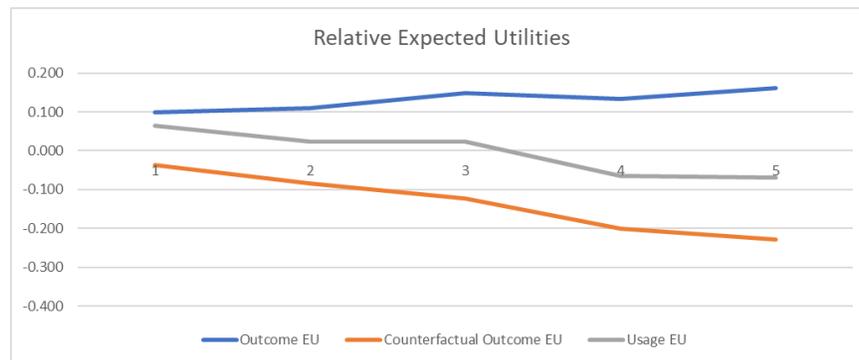

Figure 5: Sample run where DM accepts confident AI positive readings but otherwise reviews AI recommendation and explanation before making a final judgment.

Please note the following.

First, outcome-EU is higher than in Sim3 and increases with greater algorithm complementarity. Having the DM attend to explainable-AI has clear and substantial outcome benefit.

Second, counter-EU is negative at all complementarity levels and becomes more negative with greater complementarity. The previous simulation, Sim3, demonstrated how asymmetric triage successfully avoided negative counter-EU and therefore ensured a positive usage-EU. But Sim4 replaced DM-U with explainable-AI in non-triaged cases and that causes counter-EU to be negative which in turn lowers usage-EU.

The harm introduced by explainable-AI can be attributed to cases where the AI is wrong, and the explanation pushes the DM into accepting the wrong AI choice. Explainable-AI can cause usage-EU to go negative despite clear outcome benefits.

<u>Sim5: Explainable AI with less costly counterfactual misses.</u>

Costly CFNs drive the negative usage-EU results in the above analyses; and in our running example that cost is a consequence of the clinician being accused of violating medical ethics by deferring final judgment to the AI. It could be argued that with explainable-AI clinicians are *not* deferring judgment. Rather they are simply using the AI as an information source. Consequently, the cost of counterfactual misses with explainable-AI is not nearly as negative as posited above.

Assume for the moment that this argument is correct. Figure 6 shows simulation results where all parameter settings are exactly as they were in Sim5 except that the cost of a CFN is cut in half: -15 instead of -30.

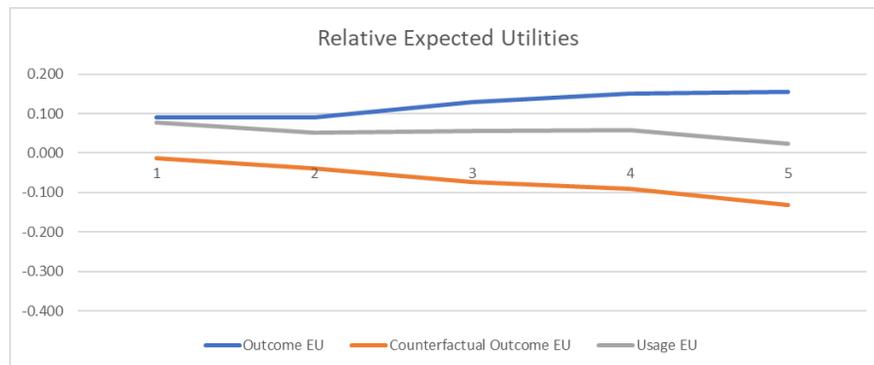

Figure 6: Sample run with explainable-AI where the cost of CFNs is reduced.

This looks quite promising. The impact of CFNs are muted and the outcome benefits of explainable-AI outweigh the cost of the CFNs that exist because the DM attended to AI explanations.

Note that the difference between Sim4 and Sim5 is entirely attributable to anticipated societal reactions. Explainable-AI provides a cover story allowing DMs to claim that when a CFN occurs they did not defer to the AI. Explainable AI is deemed more socially acceptable than automated AI and that is reflected in the low disutility for CFNs.

Unfortunately, for AI image reading, this cover story may not hold. Recall that CFNs are cases where clinical care reviews *already* concluded that the clinician would have gotten it right if she had not used the AI. The notion of "defer" is tricky here. The clinical care review will correctly conclude that if she had paid less attention to the AI recommendation, or worked harder to make an independent judgment, then she would likely have made a better decision. A post hoc judgment of having deferred to the AI is still very possible.

Sim6: Using AI as a second opinion

Finally we show results for a use pattern where it is much easier to claim a cover story that mitigates CFN costs. Specifically, using AI for a second opinion.

In use-pattern 5 (UP5)
- The DM first works each case independently.
    - If the DM confidently concludes "T" she goes with that determination.
    - If the DM confidently concludes "F" she goes with that determination.

- If the DM is uncertain, she invokes the AI and attends to the AI recommendation and explanation.

Figure 7 shows results where utilities are set as Sim5 and DM confidence thresholds are set to be the same symmetric thresholds for the AI in Sim2.

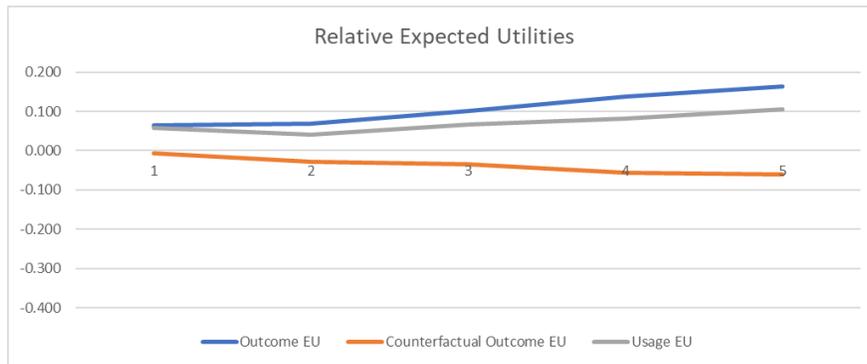

Figure 7: Sample run where DM makes initial determination and invokes explainable AI if initial determination is uncertain.

These results are very encouraging. Both outcome-EU and usage-EU improve with increasing complementarity as the impact of counter-EU is muted.

It is worth noting that this use pattern increases user workload since all cases are initially solved without AI and then AI is sometimes invoked to support additional analysis. Not modeled here is whether the additional costs outweigh the benefits of using AI in this way.

**Discussion**

*Summarizing the approach and results.*

The expected utility of deploying an AI capability (usage-EU) is the sum of the expected utility of better outcomes (outcome-EU) and the expected utility attributable to a new decision process that includes AI (counter-EU). Counterfactual misses (poor decisions attributed to using the AI) may have substantial costs. Counterfactual hits (good decisions that are attributed to using the AI) yield lesser benefit. Particularly key is the extent to which counterfactual hits and misses are noticed (discovery probabilities). Poor decisions often lead to retrospective reviews where the attribution to using the AI is discovered. Good decisions rarely lead to retrospectives, so the benefits attributable to a counterfactual hit rarely obtain. As a result, counter-EU is usually negative and that often causes usage-EU to be negative. Despite better outcomes, a rational decision maker is often wise to avoid deploying AI.

The simulation tool is designed to help developers to navigate the complex interactions between outcome-EU and counter-EU to ensure an AI system has a positive usage-EU. The simulation results showed some useful general results.

- Usage-EU is very dependent on use pattern.  A good AI algorithm can lead to either positive or negative usage-EU depending on how it is used.  For example, adding explainable-AI to a use pattern where the AI is already effectively being used for triage can substantially reduce usage-EU.  Before devoting substantial resources into explainable-AI, developers should first confirm that adding such explanations are not counterproductive to adoption.
- Characteristics of an AI algorithm that enhance outcome-EU may at the same time harm usage-EU.  For example, high algorithm complementarity where the AI and unaided DM often make different choices usually improves outcome-EU.  However, that same complementarity can make counter-EU substantially worse.  The net effect is that application contexts where the outcomes benefits of the AI are the highest may also be application contexts where usage-EU is lowest.

Figure 8 may help to visualize an explanation for many of the simulation results.

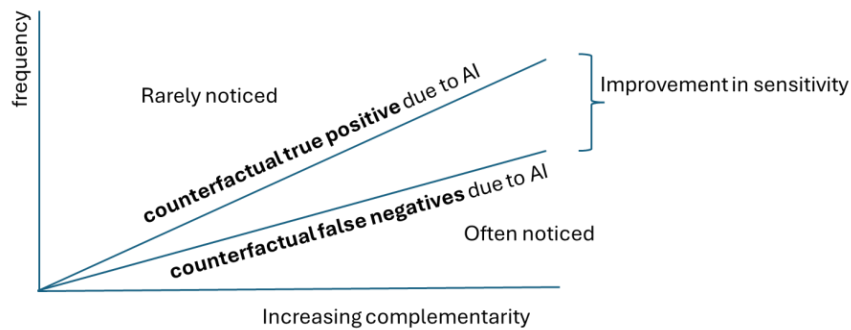

Figure 8.  Characterizing impact of complementarity on sensitivity

For most use patterns increasing complementarity leads to more CTPs and CFNs.  As long as the rate of increase for CTPs is greater than for CFNs, sensitivity will improve with greater complementarity.  However, CTPs are rarely noticed so they provide no gains beyond the better decisions.  By contrast, CFNs are often noticed and carry costs associated with these new false negatives.  There is an identical chart for specificity.

*Potential Applications*

Usage-EU analyses can be applied to examining AI adoption in diverse application areas.  Counterfactual utilities and discovery probabilities will differ and that will lead to differing conclusions about usage-EU.  To illustrate this diversity, several application areas are briefly discussed below.

> <u>Scientific discovery</u>.  Applications of machine learning to scientific discovery are growing rapidly (Wang, 2023).  Applications in areas such as material science, protein discovery, genetics, biochemistry, etc. use machine learning to generate possible solutions that are then tested or otherwise definitively vetted.  Where such vetting is routine, AI applications will not suffer from problems with counterfactual misses.  Here a CFP would be an AI-generated false discovery, but definitive vetting will quickly discard it.  CFNs would be discoveries that are missed because AI was employed rather than another scientific investigation processes. But missed discoveries have 0.0 discovery probability because, well, they were missed. From a usage-EU perspective there is no impediment to continued and growing usage in these areas.

Jurisprudence. As described earlier there are algorithms, such as COMPAS, that are widely used to support sentencing and parole decisions despite serious concerns about the fairness and accuracy of these algorithms. No doubt there are numerous cases of counterfactual misses where a COMPAS recidivism prediction led to a decision to parole someone who later commits a serious, headline-worthy, crime. It would seem that the wide adoption of a (potentially) faulty algorithm is a counterexample to the usage-EU approach described here. However, parole and sentencing decision processes are not so transparent as to separate out what decision would have been made if COMPAS had not been used. So, while potentially costly counterfactual misses do occur, they are not discoverable. Counter-EU is therefore effectively zero.

By contrast, as discussed earlier, a new and better algorithm that replaces COMPAS would be very susceptible to discoverable costly counterfactual misses. This is because it is straightforward to calculate exactly what an older algorithm would have recommended. We are currently involved in research exploring usage-EU analysis in this area.

AI symptom checkers. There is considerable interest in using large language models (e.g., ChatGPT) to provide the public with symptom checkers (Singhal, 2023). Symptom checkers would help individuals assess their symptoms, possible causes, and make recommendations as to whether they should see a medical professional. Counterfactual misses, in particular CFNs, are likely to be a problem. These would be instances where the symptom checker recommends not seeing a medical professional when, in retrospect, they should have. The symptom checker may have made the correct recommendation given the data presented, but any publicity around such cases may not take that into account. A usage-EU analysis that could examine this area might be used to assess the extent of this problem, and to find use-patterns that might avoid the worst consequences of counterfactual misses. We are currently exploring this area with a company that is fielding a medical Chatbot.

Emergency management. Machine-learned algorithms are being developed to help emergency managers to better deploy resources during natural disasters. For example, algorithms that deploy drones to efficiently discover regions with the greatest needs (Thomas, 2023). Here a counterfactual miss might be an AI-attributable decision not to allocate resources in an area later deemed in great need, perhaps leading to loss of life. While the AI resource allocation may indeed be optimal, that optimality may not protect emergency managers from the political consequences of such counterfactual misses, which in turn could lead to disuse. We are currently working with a group of state emergency managers to explore this possibility.

Opportunistic screening. This is a radiology technique that uses existing imaging data to detect health issues or risk factors that were not the original purpose of the scan (Eltorai, 2023). While clinician attention is drawn to certain elements of the image, the AI explores the rest. The clinician works DM-U. All AI finds are either CTPs or CFPs. This is an interesting use of AI image reading that initially appears to avoid the CFNs that were so problematic in the above examples and analyses. However, CFNs could easily resurface. No matter the threshold settings for detecting health issues, there will be missed diagnoses that could have been found with a lower threshold. A clinical care review might determine that the thresholds should be lowered. But lower thresholds in turn will create more CFPs, which in turn could discourage usage. A usage-EU analysis might help to determine if there are threshold settings where usage-EU is robustly positive.

Policy applications. In our running example one of the causes of negative counter-EUs is medical ethics. The policy of requiring clinicians to make a final determination tends to increase the frequency and cost of CFNs. As a result, clearly beneficial AI capabilities have a negative usage-EU. The usage-EU analysis proposed herein helps to make clear the fact that a policy intended to benefit patients may in fact be harmful. And a usage-EU analysis might help assess the trade-off of alternative ethical policies. For example, a policy where the clinician is responsible for ensuring a sound decision process, whether or not she makes the final decisions.

Robotaxis. Given the GM Cruise taxis incident, in the social context at the time of this writing, robotaxis appear to be very vulnerable to counterfactual misses where the robotaxis cause "freak" accidents. From a usage-EU perspective, there appear to be two paths to reducing counter-EU. The first is to ensure that robotaxis are almost entirely accident free. No accidents imply no counterfactual misses. Problem solved. No doubt at some future time this will be achieved. Until then, a second approach might be to highlight counterfactual hits, drawing public attention to serious accidents that would not have occurred were a robotaxi in use. It would be interesting to apply the usage-EU equation to help forge a publicity campaign favoring robotaxis.

*Future work*

Our immediate focus will be on exploring application potential. Specifically,
- Explore multiple application areas. We have started to work with other groups to explore the applicability of usage-EU analysis in three areas: emergency resource allocation, new predictive algorithms for recidivism forecasts and medical Chatbots. Feedback from these application efforts will help refine usage-EU analysis strategies and better scope broadness of relevance.[9]

---

[9] This research is funded as part of the AI institute for Societal Decision Making (https://www.cmu.edu/ai-sdm/). Our ability to quickly move toward applications is enabled by the other researcher groups in the AI-SDM already working in each of these application areas. More will follow.

- Integration into comprehensive AI adoption analysis. As noted earlier counterfactual considerations are only a part of any analysis of AI adoption. We look to combine counterfactual and traditional elements into a combined AI-adoption assessment service.[10]

Longer term there is an obvious need for further theory development. Everything presented above was for binary outcomes and decisions. Consideration of more complex decision and outcome spaces will not likely change the general form of the usage-EU equation (outcome-EU + counter-EU) but there is still much to work out.

Finally, although presented as a rational choice approach, we do not offer a formal axiomatic derivation that proves that usage-EU is the correct rational choice equation. Developing an axiomatic derivation might be an interesting exercise in that it involves assigning utilities to choices not made and probabilities to outcomes that can't occur.

---

[10] This research will be enabled by both the AI-SDM and our close working relationship with the MITRE AI Assurance Lab (https://www.mitre.org/news-insights/fact-sheet/ai-assurance-discovery-lab).

## Appendix: Simulation parameters

Below we describe the parameter values that need to be specified. We briefly describe each parameter and its import. Readers interested in detail should review the Excel spreadsheet/model and additional explanations that will be made available.

*Modeling the problem domain.*

The next three parameters characterize the problem domain.

> Prior: Sets the base rate for positive instances (GT = T)

Each simulated case is assigned a base signal strength (BSS) on a 0 to 1 scale. BSS is defined as a best possible reading of all evidence that is available to both DM and AI. The impact of evidence that is available to the DM but not the AI and vice versa is set in different parameters.

> Obviousness: This parameter sets the average distance of BSS between positive and negative readings. If obviousness is set to .4, then the mean BSS for positive instances is .7 and for negative instances is .3.

The extent to which cases are easy or difficult to solve can substantially impact AI usage utility. In easy domains, where obviousness is high, AI and DM will almost always come to the same conclusions. Here the primary value of the AI may be to increase throughput. In difficult domains, the primary benefit of the AI may be to help DMs to come to a better decision.

> Base strength standard deviation: The final BSS for each case is set to the initial BSS plus normalized variance. This parameter sets the standard deviation for the normal distribution. (Before truncation to 0 or 1.)

BSS represents evidential strength/signal that both AI and DM can perceive if both read the common evidence exactly correctly.

*Modeling AI and DM judgments.*

The next three parameters set the extent to which the AI and DM diverge in their processing of common evidence.

> AI BSS standard deviation: Additional normalized variance is added to the BSS to assign an AI perception of the common evidence. This parameter sets the distribution.

> DM BSS standard deviation: Additional normalized variance is added to the BSS to assign a DM perception of the common evidence. This parameter sets the distribution.

<u>Algorithm Complementarity</u>: Keeping overall AI and DM accuracy constant, this variable sets the extent to which the DM and AI Base Strength judgments converge. If complementarity is set to 0 then the AI and DM process common evidence exactly the same; AI Base Strength and DM Base Strength are the same. If complementarity is set to 1, then case specific BSS adjustments remain conditionally independent. Complementarity does not impact directional strength adjustments because (as modeled here) they are already inherently conditionally independent.

Research that examines how to combine AI and DM decisions often find that the combination (AI+DM) is more accurate than either alone. Furthermore, as noted earlier, they generally find that increasing levels of complementarity yield increasing accuracy. The explanation for this result is simple. If the AI and DM both make the same choices, then there is no room to improve accuracy. If they make different choices, then the proportion of cases where at least one is correct increases. Accuracy increases if the use pattern usually leads to selecting the one that is correct.

Consider a simple numerical example. Both AI and DM are 70% accurate and complementarity = 0. Combined, AI+DM accuracy will be 70%. But if complementarity = 1, then at least one of them will be correct on 91% of the cases. So AI+DM accuracy should be between 70% and 91%, depending on how effective the use pattern is in selecting the one that is correct.

The next two parameters account for information available to AI that is not available to DM.

<u>AI directional strength</u>: Beginning with the AI-BSS this specifies an average extent to which the AI final strength (AI-fs) moves in the direction of ground truth. For example, GT=F and AI-BSS is .55, and AI directional strength is .1, then the expected strength for AI-fs is .55 - .1 = .45.

<u>AI directional standard deviation</u>: This sets the extent of random normalized variance to the AI directional strength.

The next two parameters account for information available to DM that is not available to AI.

<u>DM directional strength</u>: Beginning with the DM BSS this specifies an average extent to which the DM final strength (DM-fs) moves in the direction of ground truth. For example, GT=T and DM-BSS is .55, and DM directional strength is .1, then the expected strength for DM-fs is .55 + .1 = .65

<u>DM directional standard deviation</u>: This sets the extent of random normalized variance to the DM directional strength.

*Modeling how the AI and DM interact.*

These first three parameters describe how the DM interacts with the AI to generate a final judgment.

Use Pattern: A use pattern describes a sequence of interactions between a user and an AI tool. At the time of this writing a simulation user can specify one of five different use patterns:
1. *Accept AI*: Whatever the AI recommends, the user accepts it without further consideration.
2. *Accept AI if AI confident else ignore AI*: Whenever the AI advice exceeds preset confidence thresholds that advice is routinely accepted. Otherwise the user completely ignores the AI and proceeds to work the problem herself.
3. *Accept AI if AI confident, else consider AI explanations before making judgment*. Here the user considers the AI advice even if the AI lacks confidence. Importantly, the user examines the output of the AI explanatory capability.
4. *Accept AI if AI confident, else solve independently and then review AI advice*. Here the user does not examine the AI advice or explanation until after she has independently made her own assessment.
5. *DM solves first and if not confident invokes AI*. Here the user first generates her own conclusion. If she is confident in that conclusion then she ignores the AI. If she is not confident, she invokes the AI and considers its recommendation and explanations.

Depending on parameter settings a use pattern can characterize a variety of different usage conditions. For example, in both the popular press and the AI academic literature, there is an oft assumed usage – the user reviews the AI recommendation and explanation and then makes a decision. This is a special case of UP3 where the acceptance thresholds are set to extremes to ensure no automated acceptances.

AI-acceptance thresholds. UP2 – UP4 all begin with "Accept AI if confident, else …" AI-acceptance thresholds define the "if confident" condition. Separate thresholds are assigned to positive and negative readings, so a simulation user may specify a weak threshold for accepting a positive reading (allowing more AI-attributable false positives) while specifying a stringent threshold for negative readings (thereby avoiding AI-attributable missed positives)

DM-acceptance thresholds. UP5 has the DM initially working the case and then invoking the AI if she lacks confidence in her assessment. As with AI-acceptance, the simulation user can set DM-acceptance threshold

*Modeling AI+DM combined performance.*

For UP1 and UP2, the use pattern determines whether the AI or DM judgments are selected to be the final judgment. There is no collaboration between the two. For UP3, UP4 and UP5 the DM will, in selected cases, review the AI recommendation and explanation. The following parameters model the impact of this collaboration.

> Anchor weight: This parameter only applies to cases where the DM first examines the AI recommendation before making her own final judgments. The final judgment will be weighted toward the AI recommendation.

Anchor weight reflects a robust human decision making heuristic known as Anchoring and Adjustment. People anchor on an initial judgment and then under adjust based on additional evidence. Anchor weight only applies to cases where the DM saw the AI recommendation before making her judgment. This only happens in UP3. In UP4 and UP5 the DM works the problem independently before reviewing the AI recommendation. Here the weight is set to a default of .5. That is, the AI and DM judgments are equally weighted.

It is important to note that no matter the weight, *any* linear combination of two independent judgments will on average be more accurate than either judgment alone. Consequently, the combined AI+DM judgment will on average be better than either alone even if the DM only looks at the AI recommendation and never considers the explanation.

The weighted combination of AI and DM judgments are the starting point for the final AI+DM judgment. The following parameters yield final adjustments.

> Directional Discrimination. This parameter only applies in use patterns where both AI and DM individually generate judgments and then combine those judgments. After averaging the AI and DM-U judgments, the combined judgment is shifted on average in the direction of ground truth per this parameter. For example, for UP5, if the final AI and DM-U judgments are 0.7 and 0.4, directional discrimination is .04, and ground truth is T, then the final AI+DM judgment will be (0.7+0.4)/2 + .04 = 0.59.

> Directional Discrimination std. This adds normalized random variance to directional discrimination with this standard deviation.

Directional discrimination roughly summarizes the ability to sort out the correct answer in cases where the AI and DM individually disagree. This discrimination may be based on examining the AI explanatory outputs or context information that helps to sort out circumstances where the AI or the DM tend to be more correct.

Importantly, although on average the directional adjustments are in the correct direction, this adjustment occasionally can push the final AI+DM judgment in the wrong direction.

> Explanatory boost: This parameter only applies to UP3 to cases where the AI generates a recommendation and the DM considers that recommendation, and corresponding explanation, to make a final judgment. This is modeled as an average boost in the direction of ground truth with variance in a normal distribution.

> Explanatory boost std: This sets the variance.

As with directional discrimination, explanatory boost will sometimes push the final judgment in the wrong direction.

*Utilities and discovery probabilities.*

> <u>Outcome utilities</u>. These are utilities assigned to any outcomes that are independent of the decision process used. For example, in medical applications this might be the utility associated with correct and incorrect diagnoses.
>
> <u>Counterfactual outcome utilities</u>. These are utilities assigned to counterfactual outcomes where how a decision is made impacts utility obtained. For example, a missed diagnosis attributable to using the AI may have negative repercussions well beyond the normal disutility assigned to a missed diagnosis.

In the current simulation there are two different counterfactual outcome utility matrices depending on the extent to which the AI recommendation is accepted without review or whether the AI output serves as an input to the DM. In the medical domain a poor diagnosis that resulted from simply accepting the AI output might be deemed a greater ethical breach than if the clinician reviewed the AI output to make a final diagnosis.

> <u>Counterfactual discovery probabilities</u>. This is exactly as described in the main body.

*Workload*

Finally, simulation users can specify the workload involved in different case-specific use patterns. The workload associated with simply accepting the AI output will be far less than, say, workload associated with both the AI and DM independently solving the cases and then the DM combines their judgments. Workload was not part of any of the results presented in the main body.